\documentclass[runningheads]{llncs}
\usepackage{graphicx}

\usepackage{booktabs}
\usepackage{amsmath}

\usepackage{enumitem, array}
\usepackage{xspace}
\usepackage{xcolor}
\usepackage{csquotes}
\usepackage{comment}
\usepackage{hyperref}
\usepackage[capitalise, noabbrev]{cleveref}
\usepackage{subcaption}
\usepackage{multirow}
\usepackage{colortbl}
\usepackage{listings}
\usepackage{color-edits}
\addauthor{sw}{orange}

\newcommand\newsubcap[1]{\phantomcaption%
       \caption*{\figurename~\thefigure(\thesubfigure): #1}}
       
\newcommand{\paragraphBold}[1]{\paragraph{\emph{\textbf{#1}}}}

\newcommand{\sysname}{\textsc{HypoCompass}\xspace}
\newcommand{\supplementurl}{\url{http://tinyurl.com/hypocompass-sup}}
\newcommand{\supplement}{\href{http://tinyurl.com/hypocompass-sup}{Supplements}}
\newcommand{\eg}{\emph{e.g.,}\xspace}
\newcommand{\ie}{\emph{i.e.,}\xspace}

\newcolumntype{L}[1]{>{\raggedright\let\newline\\\arraybackslash\hspace{0pt}}m{#1}}
\newcolumntype{C}[1]{>{\centering\let\newline\\\arraybackslash\hspace{0pt}}m{#1}}
\newcolumntype{R}[1]{>{\raggedleft\let\newline\\\arraybackslash\hspace{0pt}}m{#1}}

\definecolor{cprompt}{HTML}{3c4043}
\definecolor{cinput}{HTML}{3182bd}
\definecolor{coutput}{HTML}{137333}
\newcommand{\tinput}[1]{{\color{cinput}\{#1\}}\xspace}
\newcommand{\toutput}[1]{{\color{coutput}\{#1\}}\xspace}
\newcommand{\tprompt}[1]{\texttt{\textcolor{cprompt}{#1}}\xspace}
\newcommand{\colbox}[2]{\colorbox{#1}{#2}}

\newcommand{\veryshortarrow}[1][3pt]{\mathrel{%
   \hbox{\rule[\dimexpr\fontdimen22\textfont2-.2pt\relax]{#1}{.4pt}}%
   \mkern-4mu\hbox{\usefont{U}{lasy}{m}{n}\symbol{41}}}}
  
\definecolor{caddback}{rgb}{0.90, 0.98, 0.96}
\definecolor{cadd}{rgb}{0, 0.47, 0.34}
\definecolor{cdelback}{rgb}{1, 0.94, 0.92}
\definecolor{cdel}{rgb}{0.83, 0.32, 0.16}
\def \arrow{$\veryshortarrow$}
\newcommand{\add}[1]{\colbox{caddback}{\color{cadd}#1\xspace}} %

\newcommand{\remove}[1]{\colbox{cdelback}{{\color{cdel}#1\xspace}}}%
\newcommand{\swap}[2]{\remove{#1}~\arrow~\add{#2}}

\definecolor{cquote}{HTML}{3c4043}
\newcommand{\quoteinline}[1]{{\color{cquote}\emph{``#1'’}\xspace}}

\begin{document}
\title{How to Teach Programming in the AI Era? Using LLMs as a Teachable Agent for Debugging}
% \thanks{Supported by organization x.}}
%
\titlerunning{\sysname: LLM-based Hypothesis Construction Tutor}
% \author{Anonymous Author(s)}\institute{}

\author{Qianou Ma\inst{1}
% \orcidID{0009-0002-8634-130X} 
\and
Hua Shen\inst{2}
% \orcidID{0000-0002-4928-525X} 
\and
Kenneth Koedinger\inst{1}
% \orcidID{0000-0002-5850-4768} 
\and
Sherry Tongshuang Wu \inst{1}
% \orcidID{0000-0003-1630-0588}
}
\authorrunning{Q. Ma et al.}
\institute{Carnegie Mellon University, Pittsburgh PA, USA \email{\{qianoum,krk,sherryw\}@cs.cmu.edu}\\ \and
University of Michigan, Ann Arbor MI, USA\\
\email{huashen@umich.edu}}
 
\maketitle              % typeset the header of the contribution
%

% \vspace{-10pt}
\begin{abstract}
Large Language Models (LLMs) now excel at \emph{generative} skills and can create content at impeccable speeds. However, they are imperfect and still make various mistakes.
In a Computer Science education context, as these models are widely recognized as ``AI pair programmers,'' it becomes increasingly important to train students on \emph{evaluating} and \emph{debugging} the LLM-generated code.
In this work, we introduce \sysname, a novel system to facilitate deliberate practice on debugging, where human novices play the role of Teaching Assistants and help LLM-powered teachable agents debug code.
We enable effective task delegation between students and LLMs in this learning-by-teaching environment: students focus on \emph{hypothesizing the cause of code errors}, while adjacent skills like code completion are offloaded to LLM-agents. Our evaluations demonstrate that \sysname generates high-quality training materials (\eg bugs and fixes), outperforming human counterparts fourfold in efficiency, and significantly improves student performance on debugging by 12\% in the pre-to-post test.

\keywords{LLM \and teachable agent \and debugging \and CS1.}
\end{abstract}
%

% \section{First Section}
% \subsection{A Subsection Sample}
% \subsubsection{Sample Heading (Third Level)} 
% \paragraph{Sample Heading (Fourth Level)}

\section{Introduction}
\label{sec:intro}

LLMs are becoming an integral part of software development --- commercialized tools like GitHub Copilot are now advertised as ``your AI pair programmer'' and generate up to 46\% of users' code~\cite{dohmke2023copilot}.
Despite their prevalence, LLMs often produce unpredictable mistakes~\cite{ganguli2022predictability}, \eg GPT-4 can still make mistakes 17\% of the time in coding tasks for introductory and intermediate programming courses~\cite{savelka2023thrilled}. 
The impressive yet imperfect generative capabilities of LLMs, coupled with the associated risks of excessive reliance on these models, underscore the importance of teaching \emph{evaluation} skills to students.
In the context of programming, students must improve their debugging and testing skills~\cite{becker2023programming}.

However, debugging tends to be overlooked in formal educational curricula, especially in introductory Computer Science classes (\ie CS1)~\cite{whynodebugging2014}.
Prior research has outlined various factors contributing to the absence of debugging instruction, such as instructors' limited time budget for developing specialized debugging materials and assessments~\cite{McCauley2008review}.
Consequently, students primarily learn debugging from working on their own mistakes, which can be rather frustrating --- they must invest substantial time and effort in \emph{hypothesizing} the cause of bugs while grappling with other cognitively demanding tasks, such as understanding and writing code.
These challenges prompt us to ask: \\
\textbf{Research Question:}
{Can we train students to improve debugging skills by providing \emph{explicit} and \emph{scaffolded} practice \emph{with minimal cost to instructor time?}}

\begin{figure}[t]
\vspace{-1mm}
    \centering
\includegraphics[trim={0cm 5cm 0.7cm 0cm}, clip,width=.95\linewidth]{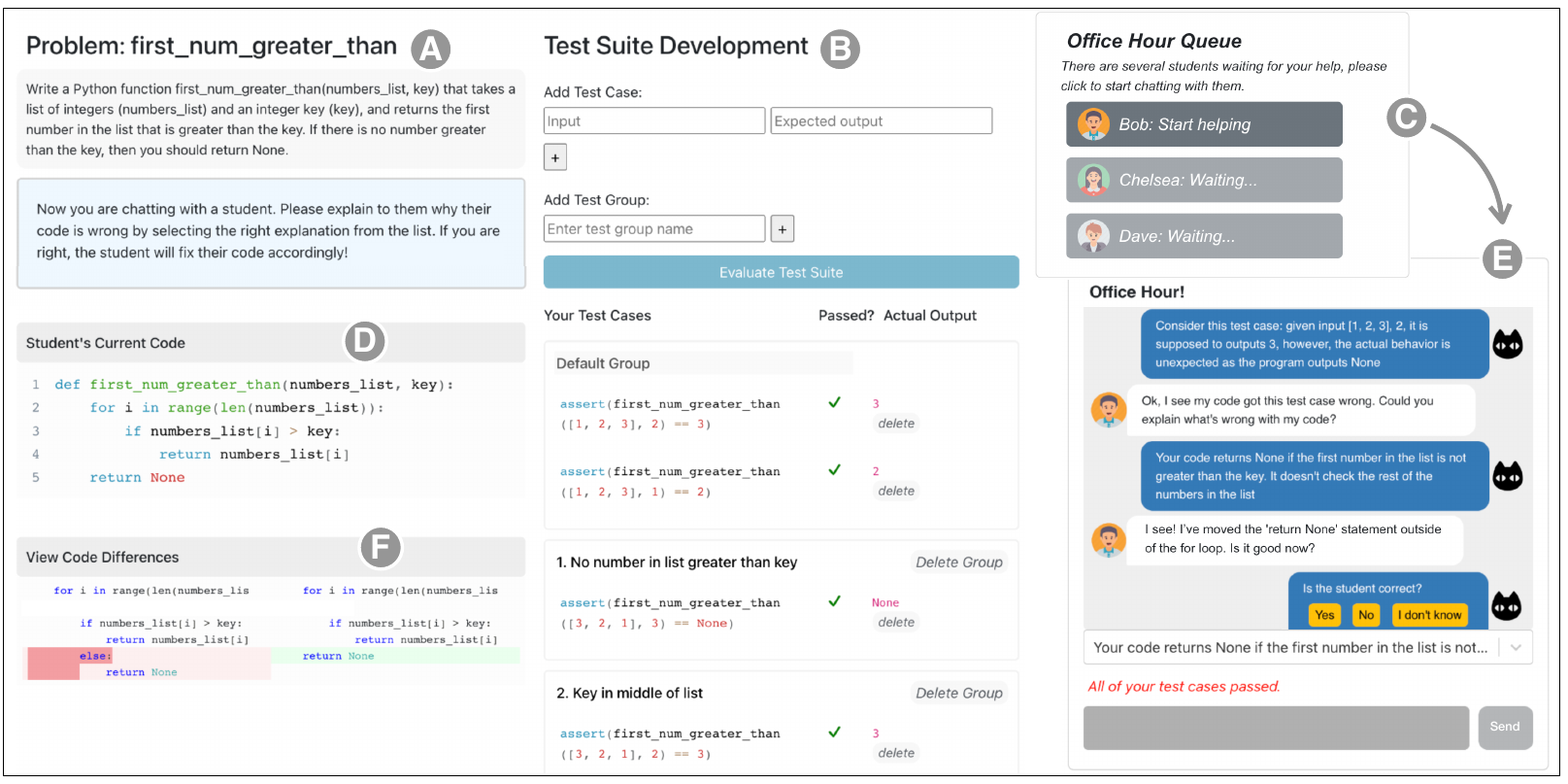}
\vspace{-10pt}
  \caption{
  In \sysname, given a programming problem description (A), 
  a student user (in the role of a Teaching Assistant) needs to compile a test suite (B) and assist multiple LLM-simulated agents (\eg \textit{Bob, Chelsea, Dave}) in an Office Hour Queue (C) through a chat interface (E). 
  Each LLM-agent acts as a novice seeking help with a buggy solution (D) and provides feedback to the user (F).
  }
  \label{fig:ui}
  \vspace{-10pt}
\end{figure}

In this work, we focus on training students' abilities in \emph{hypothesis construction}, a critical step in debugging as established by prior work \cite{Xu2004cognitive,Zeller2009why}. We introduce \sysname (\cref{fig:ui}, \cref{sec:design}), {an interactive, LLM-augmented intelligent tutoring system for debugging}. 
Leveraging LLMs' \emph{material generation} capability, we have these models imitate CS1 students who have written buggy code and require assistance from Teaching Assistants (TAs). Human novice students assume the role of the TA, who helps troubleshoot these bugs. This enables students to deliberately practice the skill of \emph{hypothesizing} about the defects of LLM-generated code, delegating other tasks not core to hypothesis construction (\eg code completion) to the LLM. 
As a result, \sysname fosters an engaging learning environment using the \emph{teachable agent} framework~\cite{Blair2007ta} and provides students with guided exposure to LLM-generated bugs.
We also {employ prompting strategies such as focused task formation and over-generate-then-select} to improve LLM generation quality in \sysname (\cref{sec:llm_integration}).

We conducted two evaluation studies and found that \sysname \emph{saves instructors' time in material generation} and is \emph{beneficial to student learning}.
In our LLM evaluation study (\cref{sec:llm_eval}), expert inspections on six practice problems and 145 buggy programs showed that \sysname achieved a 90\% success rate in generating and validating a complete set of materials, \emph{four times faster than human generation.}
Our learning evaluation study with 19 novices (\cref{sec:user-study}) showed that \sysname significantly improved students' pre-to-post test performance by 12\% and decreased their completion time by 14\%. 

In summary, we contribute:

\begin{itemize}[labelwidth=*,leftmargin=1.3em,align=left, topsep=0pt]
% \item We explicate the use of LLMs for enhancing learning, leveraging LLMs to (1) prepare students to engage with imperfect LLMs, (2) offload adjacent skills, and (3) generate materials effectively.

\item A pragmatic solution that balances the benefits and risks of LLMs in learning.
We use LLMs to prepare students to engage with imperfect LLMs, and we highlight the importance of \emph{role-playing} for practical LLM application and \emph{task delegation} to help students focus on essential skills.

\item A theoretically grounded instructional design to enhance debugging skills.
To the best of our knowledge, we are the first to provide aligned instruction and assessments on the hypothesis construction learning objectives, \ie forming hypotheses about the source of error, a core bottleneck in debugging \cite{Whalley2021analysis}.
\end{itemize}

\section{Related Works}
\label{sec:relate}

\textbf{The Debugging Process.}
\label{subsec:background}
Debugging is a complicated process of various cognitively demanding tasks, including understanding the code, finding bugs, and fixing bugs, with the first two considered primary bottlenecks~\cite{McCauley2008review,Whalley2021analysis}.
While many studies have attempted to improve students' code understanding~\cite{Kallia2023search}, there is limited instruction on bug finding. Researchers characterize the cognitive model of bug finding as a \emph{hypothesis construction process}, including initializing, modifying, selecting, and verifying hypotheses (\cref{fig:flow}B) \cite{Xu2004cognitive}. 
This process is challenging: prior works show that novices struggle to systematically generate comprehensive hypotheses and identify the right hypothesis, in contrast to experts~\cite{Edwards2014do,Edwards2014comparing}.
Hence, we emphasize teaching students to \emph{construct accurate hypotheses about bugs} and \emph{develop comprehensive hypotheses about potential bugs}.

\textbf{Tutors and Tools for Debugging Training.}
Prior studies~\cite{McCauley2008review} and online discussions~\cite{whynodebugging2014} indicate that teaching debugging is challenging and is rarely included in CS1 curricula, due to logistical challenges like the lack of instructional time and resources~\cite{Desai2009implications,Fitzgerald2010debugging}.
Existing tools demand instructor effort and often focus on the full debugging process, improving bug-fixing accuracy and efficiency~\cite{Ardimento2019reusing,LuxtonReilly2018ladebug}. 
In contrast, few studies emphasize accurate or comprehensive hypothesis construction (and they tend to be language-specific)~\cite{Ko2008debugging,Whalley2021analysis}. 
To fill in the gap, we design \sysname to provide \emph{deliberate practice}~\cite{ericsson2016peak} on hypothesis construction, and use \emph{the LLM generation capability to provide easily adaptable and targeted exercises} with immediate feedback.

\textbf{LLM Capabilities for CS Learning.}
LLMs can perform well in a CS1 classroom~\cite{savelka2023thrilled}, but concerns about misuse and LLM errors limit their use in education~\cite{becker2023programming}. 
Therefore, current deployments tend to focus on generating instructional materials (\eg questions~\cite{Wang2022towards}).
In our work, \sysname uses the LLM to generate inter-dependent materials in an integrated process and frame the LLM as a student asking for help \cite{Blair2007ta}, such that human novices can embrace imperfections in LLMs.
Two unique capabilities of LLMs power this: 
(1) LLMs can simulate different personas and tutoring interactions~\cite{Markel2023gpteach};
(2) LLMs make common mistakes and natural bugs similar to humans~\cite{mozannar2022reading}, which can be used as buggy code practice examples.
We adapt and develop various prompting methods~\cite{Wu2022aichains} to enhance the quality of LLM generations.

\section{The Design of \sysname}
\label{sec:design}

\begin{figure*}[t]
    \centering
\includegraphics[trim={0cm 20cm 10.5cm 0cm}, clip,width=1\linewidth]{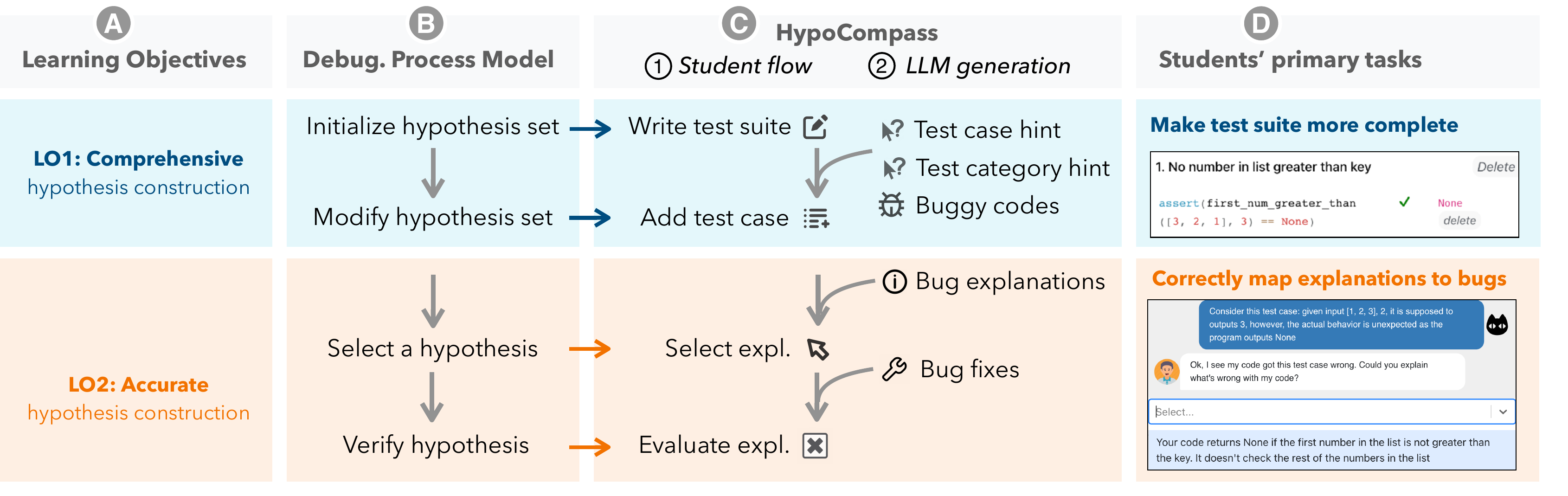}
    \vspace{-6.5mm}
  \caption{To enable deliberate practice, we establish a close mapping between the (A) learning objectives, (B) the cognitive debugging process model, (C) the \sysname interaction flow, and (D) the primary tasks students perform in \sysname. We offload various material generation tasks to LLMs (C$_2$).}%, to support the student interaction flow (C).}
  \label{fig:flow}
  \vspace{-10pt}
\end{figure*}

%We focus our design on comprehensive and accurate hypothesis construction as crucial debugging skills.
Grounded in the cognitive process \cite{Xu2004cognitive} and the novice-expert difference in hypothesis-driven debugging (\cref{sec:relate}), we specify two crucial learning components for \sysname: comprehensive and accurate hypothesis construction.
%Zeller \cite{Zeller2009why} describes this process as ``scientific debugging,'' and asserts that it can be instantiated as \emph{test suite construction}, where different types of test cases represent various hypotheses. %~\cite{Whalley2021analysis}. 
Prior work shows that hypothesis construction is closely connected with testing~\cite{Zeller2009why}:
each additional test case should, ideally, be a hypothesis about what can go wrong in the program.
In turn, a \emph{comprehensive} test suite (\ie a set of test cases) should allow an effective debugger to construct a \emph{accurate} hypothesis about why the program is wrong. 
We thus design toward two learning objectives (\cref{fig:flow}A,D):

\begin{enumerate}[labelwidth=*,leftmargin=2.6em,align=left,label=LO\arabic*, topsep=0pt]
    \item \label{lo:comprehensive} \textbf{Comprehensive Hypothesis Construction}: Construct a comprehensive test suite that well covers the possible errors for the given problem.
    \item \label{lo:accurate} \textbf{Accurate Hypothesis Construction}: Given the failed test cases, construct an accurate explanation of how the program is wrong.
\end{enumerate}

\paragraphBold{Interface and Key Components.}
%\label{subsec:interface}
We designed \sysname through an iterative development process with 10 pilots, including CS1 students, TAs, and instructors. 
%As shown in \cref{fig:ui}A, the human student is asked to play the role of a TA where they help an LLM-simulated student (LLM-agent) in debugging.
In the resulting interface (\cref{fig:ui}), a human student would be asked to play the role of a TA where they help an LLM-simulated student (LLM-agent) in debugging.
% The first step is to write a test suite that is {comprehensive} enough to catch any bug there may be in the code.
%\cref{fig:ui}A shows an example:
%\begin{displayquote}
%Write a Python function \texttt{first\_num\_greater\_than (numbers\_list, key)} that takes a list of integers (\texttt{numbers\_list}) and an integer key (\texttt{key}), and returns the first number in the list that is greater than the key. If there is no such number, return None.
%\end{displayquote}
%
% The student is asked to label groups of test cases as a step toward different \emph{hypotheses} of possible bugs. 
They need to write and sort test cases into categories (\cref{fig:ui}B) that represent different {hypotheses} of what inputs may trigger errors in code.
%\swedit{For example, test case \texttt{assert(first\_num\_greater\_than([3,2,1], 3) == None)} can fall under the category \emph{``No number in list greater than key''} in \cref{fig:hint}$_1$).}
%\sysname provides hints on initial categories, but the student is encouraged to add more.

\begin{figure*}[t]
\vspace{-5pt}
\centering
\begin{subfigure}[b]{1\linewidth}
    \centering
    \includegraphics[trim={0cm 4.5cm 28.5cm 16cm}, clip,width=0.89\linewidth]{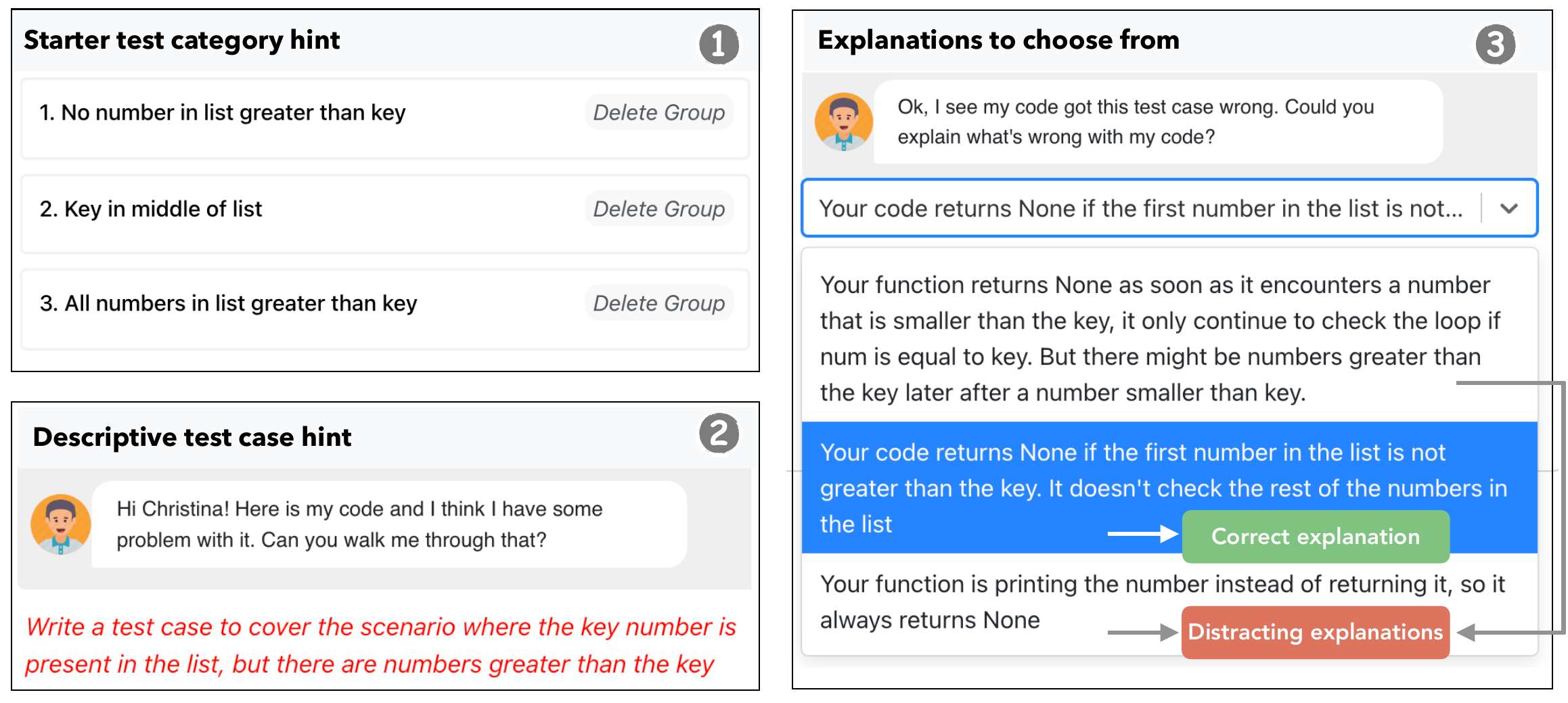}
    \vspace{-5pt} % unblock students when they struggle with the next steps, and guardrail their learning procedures
    \newsubcap{\sysname offers (1) \emph{test category hints} to help write a comprehensive test suite systematically; (2) \emph{test case hints} to help students add missing test scenarios; (3) \emph{candidate explanation pool} to clarify misconceptions of alternative explanations.}
    \label{fig:hint}
\end{subfigure}

\begin{subfigure}[b]{1\linewidth}
    \centering
    \includegraphics[trim={0cm 24.5cm 28.5cm 0cm}, clip,width=0.89\linewidth]{figures/hint_and_feedback1.pdf}
        \vspace{-5pt}
    \newsubcap{\sysname provides immediate feedback to (1) \emph{incorrect test cases}, ensuring students understand the code behavior; (2) \emph{correct explanations}, as correct code fixes; (3) \emph{incorrect explanations}, as confusion messages from the LLM-agent.}
    \label{fig:feedback}
\end{subfigure}
\vspace{-20pt}
\end{figure*}

Once the student is satisfied with their test suite, \sysname shows them an Office Hour Queue (OHQ) simulator  (\cref{fig:ui}C). %, allowing the student to {assist a series of LLM-agents}. 
As the student interacts with each LLM-agent, the agent presents a buggy code snippet (\cref{fig:ui}D). 
The student {guides the LLM-agent in debugging code through a dialog interface} (\cref{fig:ui}E), 
% constructing hypotheses on why the code malfunctions, 
selecting or creating test cases that reflect their hypotheses of the bug, and selecting explanations for the bug among a pool of candidate natural language explanations.
These candidates each explain a different bug, representing alternative hypotheses that may confuse students (\eg \cref{fig:hint}$_3$).
%\swedit{Here, the student notices that the code fails on \texttt{assert(first\_num\_greater\_than([1,2,3], 2) == 3)}. The student then selects the natural language explanation for the bug (\cref{fig:ui}E): \emph{``Your code returns None if the first number in the list is not greater than the key. It doesn't check the rest of the numbers in the list.''}}

The LLM-agent then {uses the test case and explanation to revise the code}, providing immediate feedback to the student (\cref{fig:feedback}).
If the explanation is correct, the agent will conduct minimal code fixes, and present the color-coded edits as feedback (\cref{fig:ui}F, a zoomed-in view is in \cref{fig:feedback}$_2$). 
% (here, the \texttt{return None} statement is moved outside of the loop).
Otherwise, the LLM-agent will ask the student to reflect on their hypothesis by responding with a confusion message that highlights the discrepancy between the student's explanation and the actual code behavior (\cref{fig:feedback}$_3$).

Once the student correctly confirms that all the bugs are fixed, they can {move to help the next LLM-agent} (\cref{fig:ui}C). Upon completion, \sysname will provide the next round of exercises with another programming problem. 
While the numbers are configurable, by default \sysname includes two programming exercises, each with three LLM-agents (buggy programs).

We highlight the two most essential components of the interaction:

\begin{itemize}[labelwidth=*,leftmargin=1.3em,align=left,topsep=0pt]

\item \textbf{Frame imperfect LLMs through role-play.}
We use the LLM to simulate students who wrote bugs and have human novices offer help.
This teachable agent setup supports learning, helping students reflect on their knowledge and reason through diverse bugs~\cite{shahriar2023and}. % helps TAs reinforce their problem-solving skills~\cite{roberts1995using}.
Having students work through ``other people's errors'' also boosts their motivation and protects their self-efficacy~\cite{Blair2007ta}.
More importantly, it actively involves novices in identifying bugs in LLM-generated code, \emph{enabling guided exposure to LLM imperfectness}.

\item \textbf{Task delegation between students and LLMs.}
To ensure deliberate practice on comprehensive and accurate hypothesis construction, %we ground the \sysname interaction with our learning objectives.
students primarily engage in two tasks corresponding to each learning objective (\cref{fig:flow}D): 
(1) making the test suite more complete (\ref{lo:comprehensive}); 
and (2) correctly mapping explanations to bugs (\ref{lo:accurate}).
We align student interaction flow (\cref{fig:flow}C$_1$) with the cognitive model of debugging~\cite{Xu2004cognitive} (\cref{fig:flow}B).
LLMs take over other tasks that are \emph{indirectly related} to the core learning goals, including generating diverse bugs and fixes, which frees students from code writing.
We also use LLMs to support scaffolding, generate hints (\cref{fig:hint}), and provide immediate feedback throughout the practice (\cref{fig:feedback}).%, \eg the LLM-agent reacting differently to students' correct and incorrect explanations).
%We present hints and feedback in an engaging manner, reinforcing the teachable agent role play.

\end{itemize}

\begin{figure*}[t]
\centering
    \includegraphics[trim={0cm 8cm 5cm 0cm}, clip,width=\linewidth]{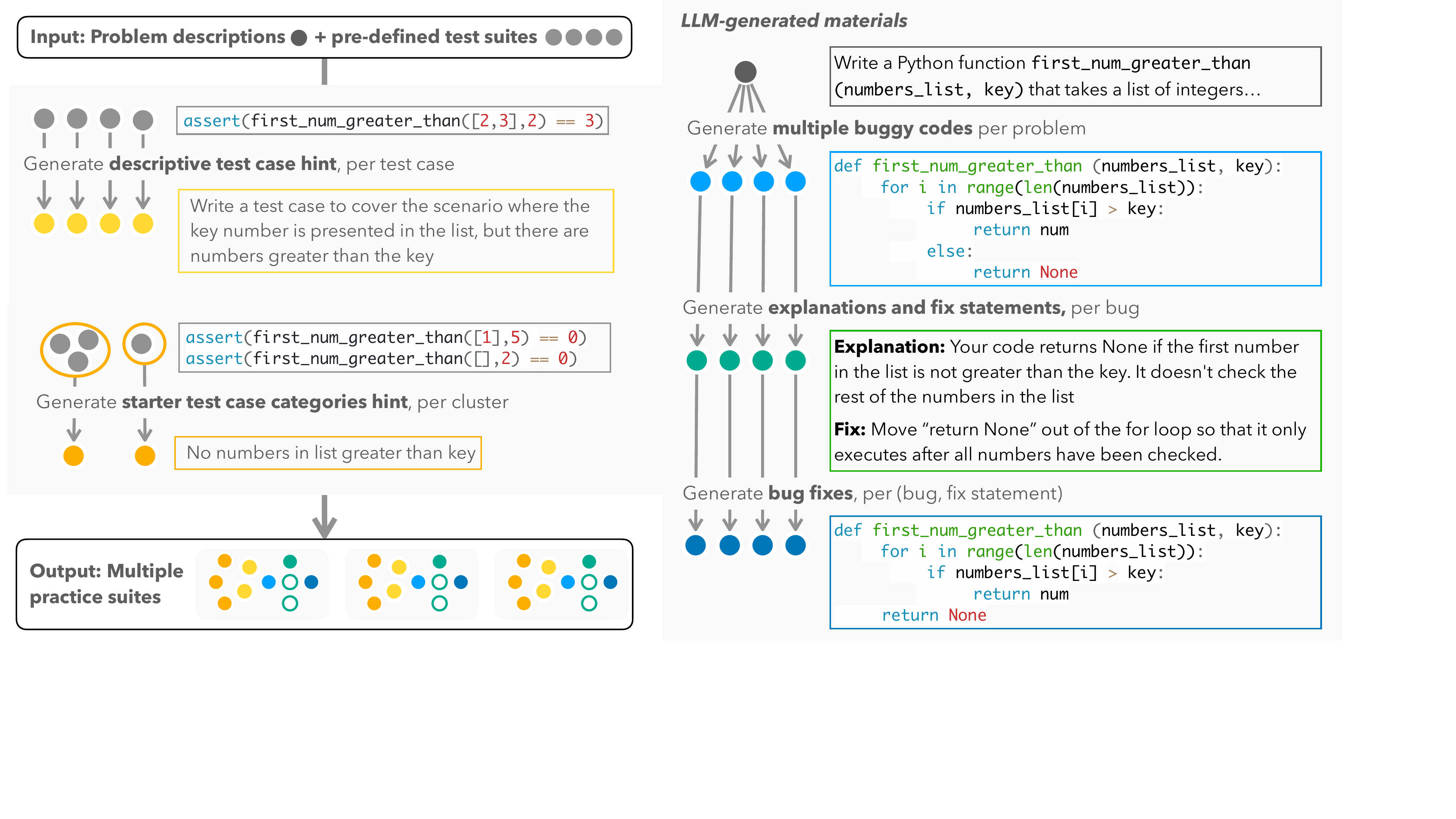}
    \vspace{-7mm}
    \caption{Examples of inputs and outputs to the LLM material generation pipeline.}
    %. Given a program exercise description and its corresponding pre-defined test suite, \sysname generates all the required training materials in an end-to-end manner. The test case and test category hints are generated independently, whereas materials relevant to bugs are generated sequentially. }
    \label{fig:llm_gen}
    \vspace{-10pt}
\end{figure*}

\section{LLM Integration}
\label{sec:llm_integration}

As shown in \cref{fig:flow}C$_2$, we use LLM to generate five types of materials: 
(1) test case category hints, 
(2) test case hints,
(3) buggy programs,
(4) explanations of bugs, and
(5) programs with bugs fixed.
We reduce instructor workload by generating practices using just a problem description, a reference solution, and a reference test suite with about 10 inputs, and we further minimize human verification overhead with optimized prompts and automated algorithms.
Our generation process is detailed in \cref{fig:llm_gen}, example prompts are in \cref{tab:prompt-ex}, and full prompts are in Table 3 in \supplement\footnote{Supplemental materials are at: \supplementurl}. %  \cref{tab:prompt} in appendix
% To balance cost and quality, we employ less advanced models in early prototypes, transitioning to more advanced ones only if smaller models consistently underperform. 
OpenAI's \texttt{gpt-3.5-turbo} is used for all materials, except for explanation generation, which uses \texttt{gpt-4} for enhanced reasoning capabilities. %Prompts and temperatures are specified in Table 1, with the top-1 output from the language model consistently chosen.
Below are key factors to the success of generation:

\paragraphBold{Task Formation and Decomposition.}
\label{subsec:task_formation}
We iterate on our prompts according to the nature of the task. 
First, as LLMs behave inconsistently when the user tasks conflict with LLMs inherent training objectives~\cite{xie2023adaptive}, we carefully {formulate the task} to avoid introducing competing tasks.
Take \emph{Local Bug Fix} (\cref{tab:prompt-ex}) as an example: when we directly ask the LLM to fix a bug according to an explanation, we observe that the model almost always over-fix all bugs irrespective of the provided instructions. 
This is because LLMs can be biased towards {generating fully correct code} (part of the LLM pre-training) and away from {local bug fixing} (changing only the buggy snippet described by the instruction, the desired task). 
Hence, we re-frame it as a \emph{translation task}, converting bug-fixing instructions to its code format \texttt{old $\rightarrow$ new code snippet}. This \textbf{task re-framing} mitigates the model's inherent bias, reducing over-fixing errors by 70\%.

Second, for multi-step tasks (\eg \emph{Local Bug Fix}), we adopt \textbf{LLM-chains}~\cite{Wu2022aichains}, decomposing tasks into sub-tasks handled by separate steps, such that each step contributes to stable performance.
Third, we also address prompt complexity by explicitly prioritizing essential requirements. For tasks like generating \emph{Bug Explanations and Fix Instructions} (\cref{tab:prompt-ex}), we prioritize precise bug extraction, instructing the model to {list all unique bugs} upfront. Secondary requirements (\eg word limits) are specified only in the output format. This \textbf{hierarchical disentanglement} significantly improves success rates by over 40\%.

\newcommand{\sysprompt}[1]{\tprompt{[Sys.]} &
\multicolumn{3}{p{0.92\textwidth}}{\tprompt{#1}} \\}
\newcommand{\userprompt}[1]{\tprompt{[User]} &
\multicolumn{3}{p{0.92\textwidth}}{\tprompt{#1}} \\}
\newcommand{\llmprompt}[1]{\tprompt{[LLM ]} &
\multicolumn{3}{p{0.92\textwidth}}{\tprompt{#1}} \\}

\newcommand{\materialtitle}[1]{ \multicolumn{2}{p{0.14\textwidth}|}{\textbf{#1}} }

\renewcommand{\arraystretch}{1}

\begin{table*}[t]
\caption{Prompts and temperatures ({Temp.}) for generating bugs, explanations, and fixes. The temperature is set higher for more diverse and random outputs.}
\fontsize{6.5}{7}\selectfont
%\small
\begin{tabular}{R{0.07\textwidth} p{0.08\textwidth} | p{0.79\textwidth} | r}
\toprule
\materialtitle{Material} & \textbf{Generation goal} & \textbf{Temp.} \\

%%%%%%%%%%%%%%%%%%%
\arrayrulecolor{black!100}\midrule 
\materialtitle{Buggy code}
     & To over-generate bugs with mixed quality for further selection.
     & 0.7 \\
\arrayrulecolor{black!30}\midrule
 %%%%%%%%%%%%%%%%%%%
 \sysprompt{You are a \textbf{novice student in intro CS}, you make mistakes and write buggy code.}
 \userprompt{
 Problem Description: \tinput{problem\_description} \newline
 Write \textbf{different buggy solutions} with common mistakes like novice students:}
 % \llmprompt{\toutput{buggy code \#1}\toutput{buggy code \#2}\toutput...}
 % \llmprompt{\toutput{buggy code \#2}}
 % \llmprompt{\toutput{...}}
%%%%%%%%%%%%%%%%%%%%

%%%%%%%%%%%%%%%%%%%%
\arrayrulecolor{black!100}\midrule
\materialtitle{Bug expl. \& fix instruct.}
    & To describe each unique bug, and write a corresponding fix instruction. If there are multiple bugs in the code, generate their explanations and fixes separately.
    & 0.3 \\
\arrayrulecolor{black!30}\midrule
%%%%%%%%%%%%%%%%%%%
\sysprompt{You are a helpful and experienced \textbf{TA} of an introductory programming class.}
\userprompt{Hi, I'm a student in your class. I'm having trouble with this problem in the programming assignment: \tinput{problem\_description} 
Here's my buggy code: \tinput{buggy\_code}
What's wrong with my code? \textbf{List all the unique bugs included, but do not make up bugs}. For each point, put in the format of: \{explanation: accurate and concise explanation of what the code does and what the bug is, for a novice, fix: how to fix the bug, within 30 words\} \newline 
Only \textbf{return the bullet list}. Do not write any other text or code.}
% \llmprompt{\toutput{Bullet list of json formats with explanation and fixes}}
%%%%%%%%%%%%%%%%%%%%

\arrayrulecolor{black!100}\midrule
\materialtitle{Bug fix}
    & To edit the buggy code according to the fix instruction, w/o over- or under- fix.
    &  0.3 \\
\arrayrulecolor{black!30}\midrule
%%%%%%%%%%%%%%%%%%%
\sysprompt{You fix bugs in Python code closely following the instructions.}
\userprompt{
Original code: \tinput{buggy\_code}; Code modification: \tinput{explanation} \newline
\textbf{Translate the statement into actual, minimal code change in this format}: \newline
\{original code snippet: ""copy the lines of code that need editing"" \newline -> edited code snippet: ""write the edited code snippet""\}}
\llmprompt{\toutput{old to new snippet in JSON, e.g., \texttt{\swap{numbers\_list[i] <= key}{numbers\_list[i] > key}} }}
% \arrayrulecolor{black!0}\midrule
% \sysprompt{You fix bugs in Python code closely following the instructions.}
\userprompt{
Old Code:\tinput{buggy\_code}; Instruction:\tinput{Old snippet to new snippet}; New Code:}
% \llmprompt{\toutput{The complete version of the new fixed code}}

%%%%%%%%%%%%%%%%%%%%

\arrayrulecolor{black!100}\bottomrule

\end{tabular}

\label{tab:prompt-ex}
\vspace{-10pt}
\end{table*}

\paragraphBold{Over-Generate-then-Select.}
\label{subsec:generate_and_select}

\begin{figure*}[t]
\vspace{-5pt}
\centering
    \includegraphics[trim={0cm 22cm 21cm 0cm}, clip,width=0.9\linewidth]{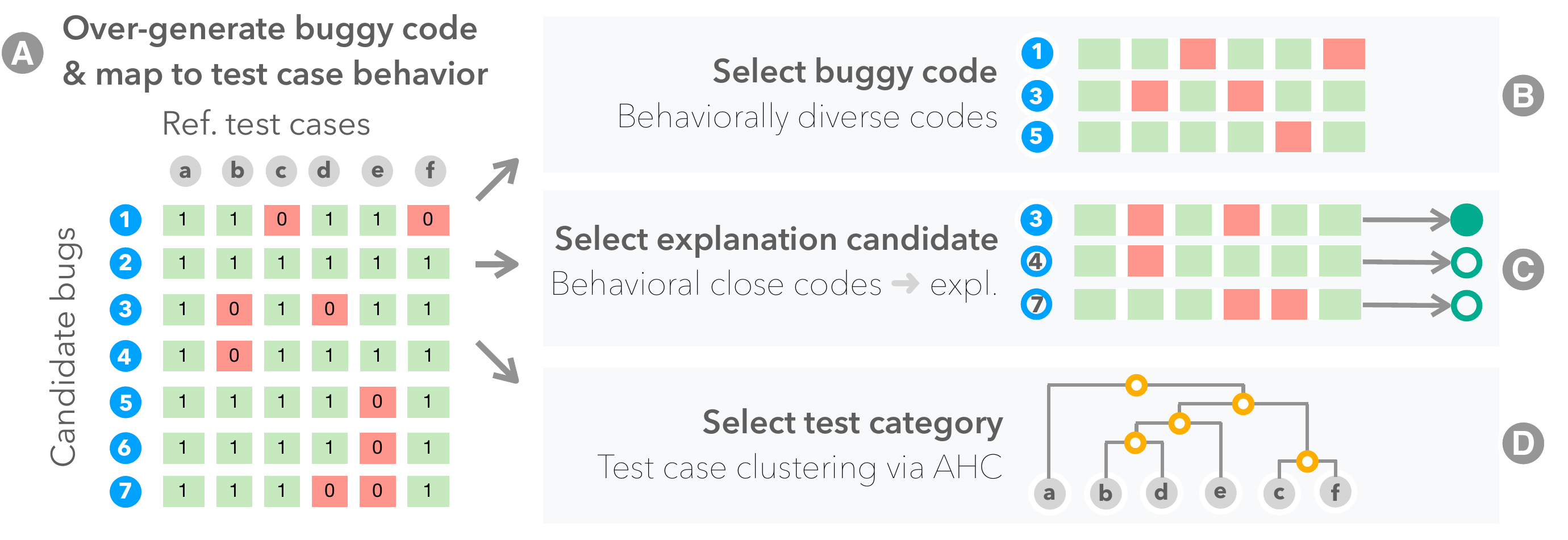}
    \vspace{-10pt}
    \caption{Over-generate and automatically select materials with pedagogical values.} %Specifically, we (A) over-generate buggy codes and calculate behavioral vectors using reference test cases. We (B) select buggy codes with distinct test behaviors, (C) select distractor codes that behave similarly for the explanation pool, and (D) cluster test cases to guide test category hints revision.}
    \label{fig:generate_and_select}
    \vspace{-10pt}
\end{figure*}

While LLMs can easily generate random materials, it is nontrivial to ensure that their generations have pedagogical values. 
For example, {behaviorally distinct} bugs help students practice with varied instances, but it is hard to enforce through prompting as it requires LLMs to ``know'' bug behaviors. Nonetheless, we can configure the non-deterministic LLMs to \textbf{over-generate} multiple solutions with mixed qualities~\cite{macneil2022generating}, and \textbf{then select a subset} of desired ones (\cref{fig:generate_and_select}). We apply this strategy in multiple places:

(1) To {expose students to behaviorally distinct bugs}, we over-generate buggy code (\cref{tab:prompt-ex}). We filter out correct code, and we vectorize buggy code's behavior based on the reference test suite (\cref{fig:generate_and_select}A, 0 being failed tests). We then greedily choose a diverse subset of buggy programs with the maximum pairwise distance, using Euclidean distance on the error vectors (\cref{fig:generate_and_select}B).

(2) To {help students clarify misconceptions} (\cref{fig:generate_and_select}C), we want distracting explanations that look similar to the actual explanation for each practice buggy code.
We choose from the over-generated buggy code pool, find two with the smallest Euclidean distance to the target code, and use their corresponding explanations as distractors. 
The mapping also helps generate the confusion messages (\cref{fig:feedback}$_3$) --- when a student selects the distractor explanation, we use its corresponding buggy code to find test cases to present to students.
% \swdelete{\footnote{We also experimented using the code syntax distances calculated with AST tree distance and RougeLCSUM \cite{Koutcheme2023evaluating}. While these methods prioritize codes with syntax differences, it is possible to have codes that look different but behave the same. We stick to the error vector distance as it allows us to focus on high-level code behaviors instead of implementation details.}}

(3) To {capture key testing aspects} in our test category hints (\cref{fig:generate_and_select}D), we cluster reference test cases into semantically meaningful groups. We build dendrograms from test case vectors with Agglomerative Hierarchical Clustering~\cite{lukasova1979hierarchical}, which guide the selection of test category hints from the over-generated pool.

\paragraphBold{Human-in-the-Loop Verification.} 
As shown in \cref{fig:llm_gen}, while the hints for test cases and categories are generated separately, the materials relevant to bugs are generated in sequential order.
We perform human verification per step to mitigate the risk of cascading errors in subsequent steps.
% For buggy code, human editing means removing extra comment lines from the generated code; while for others, it means revising incorrect snippets in the generated material or ensuring the material follows certain formats that enable subsequent parsing. 
We provide more details on human verification and editing times in \cref{sec:llm_eval}.
% As of now, all the materials in \sysname are \emph{pre-generated} in order to minimize the risk of students being misled by incorrect LLM responses.
\section{LLM Evaluation: Generation Efficiency and Quality}
\label{sec:llm_eval}

We evaluated the generations on six different problems from prior work~\cite{dakhel2023github} and our own problems (detailed in Table 4 in \supplement). % \cref{tab:problems} in appendix
On average, for each problem, we generated 3 test category hints, 10 test case hints, 24 buggy programs, explanation and fix instructions, and 33 bug fixes. The total number and the success rates are summarized in \cref{tab:llm-eval}. We provided the success criteria for all types of materials in Table 5 in \supplement. %  \cref{tab:metrics} in appendix

\paragraphBold{Method.}

Two authors annotated 10\% of the generations at each step individually, and discussed to resolve the disagreement and update the codebook. An external instructor annotated the same 10\% of LLM-generated materials, using the updated codebook. 
We calculated the inter-rater reliability (IRR) between the external instructor and the resolved annotation among the two authors using percent IRR and Cohen's Kappa. As shown in \cref{tab:llm-eval}, the agreements are satisfactory across different model generations (IRR\% $>$ 90\% and $\kappa > 0.75$)\footnote{\label{footnote:kappa}
Buggy programs undergo automatic testing, so human verification is unnecessary (n/a). If both raters unanimously agree in one category, $kappa$ is undefined (-), so $\kappa$ is only noted when there's less than 100\% IRR agreement on a single label.
}.
One author 
% \footnote{This author is an educational expert who is experienced in creating instructional materials for CS1 courses.} 
annotated the rest of the materials to calculate the success rates.
We log the verification and editing \emph{time}, as proxies to the instructor overhead.

To compare LLM and human generations, 
we recruited two experienced CS TAs to each create practice materials for a specific problem. 
Each TA received the same input as LLMs, was asked to produce one set of materials matching the amount of content LLMs produced, and was compensated for their time.

% \begin{table*}
% \Description{Description of the table.}
% \label{tab:llmeval}
% %\fontsize{7.5}{8}\selectfont
% \small
% \begin{tabular}{r | r r r | r r}
% \toprule
% \multirow{2}{*}{\textbf{Material}} &
%     \multicolumn{3}{c|}{\textbf{Raw LLM outputs}} &
%     \textbf{Human verification} \\
%     \cmidrule(){2-4}\cmidrule(){5-6}
%     & {\# Generation} & {Avg. gen time} 
%      & {Success\%}  & {Avg. edit time} & {IRR (kappa)} \\
% \midrule

% Test case description hint & 61 & 0:00:37 & 98.36\%  & 0:00:08  &  100\% \\
% Test case category hint & 18 & 0:00:10 & 94.44\%  & 0:00:10  &  100\% \\
% Buggy code & 145 & 0:01:30 & 57.93\%  & 0:00:02  & n/a  \\
% Bug explanation and fix instruction & 145 & 0:03:36 & 91.72\%  & 0:00:52  & 83\% (0.783) \\
% Bug fix & 195 & 0:02:45 & 86.15\%  & 0:00:37  &  96\% (0.780) \\
% % expl Coehn's Kappa: 0.783068783068783; Percent IRR 0.8292682926829268
% % fix Coehn's Kappa: 0.7796610169491526; Percent IRR 0.9615384615384616

% \bottomrule
% \end{tabular}
% \caption{LLM Evaluation: Time and Success rate. LLM can consistently generate high quality requirements that can be manually verified and edited in short amount of time.}
% \label{tab:llm-eval}
% \vspace{-15pt}
% \end{table*}

% =======================

\begin{table*}[t]
\vspace{-5pt}
\caption{LLM Evaluation: Time, Success rate, and Inter-Rater Reliability scores (\ie IRR\% = \#agreements / \#total labels, $\kappa$ is Cohen's Kappa coefficient).$^{\ref{footnote:kappa}}$ 
% LLM can consistently generate high-quality requirements that can be manually verified with moderate agreement and edited in a short amount of time.
}
\fontsize{7.5}{8}\selectfont
%\small

\label{tab:llm-eval}
\begin{tabular}{r | r r r | r r r}
\toprule
\multirow{2}{*}{\textbf{Material}} &
    \multicolumn{3}{c|}{\textbf{Raw LLM outputs}} &
    \multicolumn{3}{c}{\textbf{Human verification}} \\
    \cmidrule(){2-4}\cmidrule(){5-7}
    & {\# Generation} & {Avg. gen time} 
     & {Success\%}  & {Avg. edit time} & {IRR\%} & { $\kappa$ } \\
\midrule

Test case description hint & 61 & 0:00:37 & 98.36\%  & 0:00:08  & 100\% & - \\
Test case category hint & 18 & 0:00:10 & 94.44\%  & 0:00:10  &  100\% & - \\
Buggy code & 145 & 0:01:30 & 57.93\%  & 0:00:02  & n/a & n/a \\
Bug explanation and fix & 145 & 0:03:36 & 91.72\%  & 0:00:52  & 90\% & 0.875 \\
Bug fix & 195 & 0:02:45 & 86.15\%  & 0:00:37  &  92\% & 0.752 \\
% expl Coehn's Kappa: 0.783068783068783; Percent IRR 0.8292682926829268
% fix Coehn's Kappa: 0.7796610169491526; Percent IRR 0.9615384615384616
% E2 when getting started with IRR (so slower)
% explain bug: 30 min / 14
% format tc hint: 10 min / 10
% fix code round 1: 30 min / 26
% fix code round 2: 10 min / 26

\bottomrule
\end{tabular}
\vspace{-5pt}
\end{table*}

%%%%%%%%%%%%%%%%%%% metrics

\paragraphBold{Result: Efficient and High-Quality Generation.}
\label{subsubsec:llm-eval-result}
% The statistics of the number of generations, success rate, and average editing time are summarized in \cref{tab:llm-eval}.
We achieve high-quality generation: a complete set of practice materials with 9 buggy programs (3 for practice and 6 more as distractors), 9 bug explanations, 9 bug fixes, 10 test case hints, and 3 test category hints can be generated with a 90\% success rate and only takes 15 minutes to label and edit. 
As we \emph{over-generate} and automatically select buggy code, a success rate over 50\% is reasonable for practical use.

Employing LLMs can also be significantly more efficient. 
In total, a TA spent around 60 minutes to generate one set of practice materials for \sysname. 
% All the buggy codes generated by the TAs contained only a single bug, and the TAs expressed frustration during the process.
One TA noted the difficulty in consistently creating unique and high-quality materials after 30 minutes, saying that \quoteinline{the importance of the bug I create would start to decline.}
The same author evaluated the TAs' generations using the annotation codebook, which had a 100\% success rate and took 11 minutes. 
The time invested in generating and editing instructional materials for \sysname using LLMs was 4.67 times less than that of the human TAs. 

\section{Learning Evaluation: Pre- / Post-Test Study}
\label{sec:user-study}

\emph{Can novices better formulate hypotheses after engaging with \sysname?} 
We conducted a learning evaluation with 19 students and compared the difference in speed and performance from the pre-test to the post-test.

\paragraphBold{Assessment.}

% Existing methods for assessing students' debugging skills often prioritize speed and accuracy over the ability to hypothesize and identify bugs (\ref{lo:accurate}). While testing against all students' buggy code seems to assess the comprehensiveness (\ref{lo:comprehensive}) of their test suites~\cite{Edwards2014-hz, Edwards2014-ie}, this approach is not universally applicable and fails to reveal whether students generate test cases randomly or systematically.
To best capture student learning gains on our learning objectives, we took a backward design method~\cite{Wiggins2005ubd} to create an aligned assessment for the comprehensive \ref{lo:comprehensive} and accurate \ref{lo:accurate} hypothesis construction skills.
We conducted multiple rounds of pilots to refine our intervention and pre-post tests.
%, including both a multiple-choice and an open-ended version of the assessment. In the pilot tests, we observed that both formats exhibited similar completion times and accuracy rates. As such, we used the multiple-choice version to simplify evaluation.
\begin{figure*}[t]
    \centering
    \vspace{-5pt}
\includegraphics[trim={0cm 21cm 29cm 0cm}, clip,width=.9\linewidth]{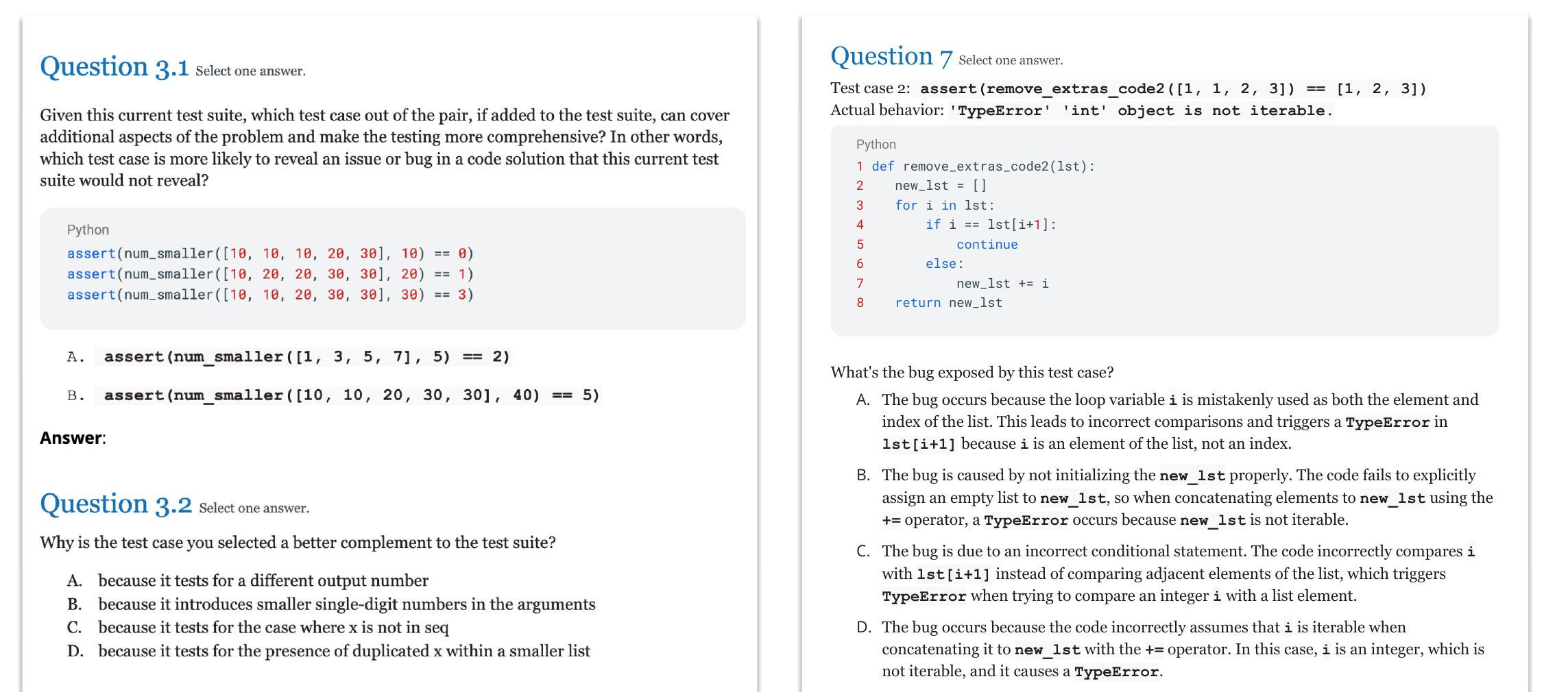}
    \vspace{-2mm}
  \caption{Pre-post test question examples for \ref{lo:comprehensive} comprehensive (Q3.1 and Q3.2) and \ref{lo:accurate} accurate hypothesis construction (Q7).}
  \label{fig:prepost-ex}
  \vspace{-10pt}
\end{figure*}
Our final tests are based on two programming exercises with comparable difficulties. We counterbalanced pre-post tests' problems to control for problem sequence influence.
Each test consists of seven questions, with three assessing \ref{lo:comprehensive} and four for \ref{lo:accurate}.
\cref{fig:prepost-ex} provides a sample for each. For instance, Question 3.1 asks students to identify the more suitable test case to add to an existing test suite, evaluating their ability to construct comprehensive hypotheses (\ref{lo:comprehensive}). 
We measure students' performance using their \textbf{test scores} based on a standard rubric. We also log the pre-post tests' \textbf{completion time} as a proxy for proficiency.

\paragraphBold{Method: Study Procedure and Participants.}

Our hour-long user study constituted a pre-survey, pre-test, interaction with \sysname, post-test, and a post-survey. 
Participants began with a pre-survey, which asked demographic information and 7-level Likert Scale questions on their debugging experiences.
Then, participants had up to 20 minutes for the pre-test. % (average completion time $12:43 \pm 4:12$).
The system interaction consisted of two problems, where participants needed to write a test suite and explain bugs in three different buggy programs for each problem.
The first problem was the same as in the pre-test, and the second problem matched the screening survey's exercise. %(\texttt{first\_num\_greater\_than}).
By reusing problems that students have seen, we isolate our learning objectives from the program comprehension skills.
After a subsequent 20-minute post-test, participants filled out a post-survey with Likert Scale and open-ended questions on their experience and perceptions using \sysname. 
% We asked whether their test cases correspond to certain hypotheses they have in mind, whether they found the code fixes reasonable, whether they think the buggy code are written by AI or student, how engaging, fun, and frustrating was the interaction, and whether they found \sysname helpful in learning about debugging.
Participants received a \$15 Gift Card for their time.

We recruited a diverse group of undergraduate and graduate students from four public or private US institutions. Interested participants completed a screening survey, which included a programming exercise 
% (\texttt{first\_num\_ greater\_than} as in \cref{fig:ui}) 
that also served as the second exercise in our study. To ensure a suitable skill range, we excluded those with extensive programming experience or who quickly solved the exercise. After filtering, 19 participants (S1-19) were included in the study --- 12 females, 6 males, 1 non-binary, and 8 non-native English speakers, with an average age of 20.7.

\paragraphBold{Quantitative Result: Learning Gains.}
A two-tailed paired t-test showed that students' pre-test to post-test scores significantly improved by 11.7\% ($p=0.033 < 0.05$), and the time of completion significantly reduced by 13.6\% ($p=0.003$), indicating success in learning through \sysname interaction.
Note that the bugs used in pre-post tests are generated by humans and are not the same as in \sysname. As such, the significant learning gains indicate that students could learn debugging skills \emph{transferable} to real-world bugs.
% I think what I've learned in this interaction will be helpful for debugging programs I wrote outside of the system. average = 6/7 Agree

Where does the learning gain come from? We break down the analyses by learning objectives.
We found a small 6.1\% improvement in the score and a large 23.6\% time reduction for \emph{comprehensive hypothesis construction} (\ref{lo:comprehensive}), and a large 15.8\% improvement in the score and a small 9.0\% time reduction for \emph{accurate hypothesis construction} (\ref{lo:accurate}). Therefore, students showed more efficiency enhancement in \ref{lo:comprehensive}, and more learning gains in \ref{lo:accurate}.
Note that these improvements may confound with problem difficulty, as the items corresponding to \ref{lo:comprehensive} (pre-test $\mu = 54\%$) seem easier than the ones for \ref{lo:accurate} (pre-test $\mu = 38\%$).

\paragraphBold{Qualitative Result: Student Perceptions.}
\label{subsec:qualitative}
We further unpack how \sysname contributed to learning by analyzing the survey responses.
Students valued being able to offload some debugging subtasks to \sysname, such as writing code and explanations. For example, S1 said \quoteinline{looking at the test behavior and the explanation options really helps relieve that burden.} %, highlighting the potential of task delegation.
% For example, three participants found it very helpful to have explanation options provided. U1 said \quoteinline{I feel such a relief that I don't need to debug myself, even when I look at the code and think it looks right, looking at the test behavior and the explanation options really helps relieve that burden.}
% P4 in the pilot study added, \quoteinline{I can just leave the Python debug part to it, so I can just find the problem and evaluate the fix automatically instead of climbing on a whiteboard to wonder and process all the info.}
% We also received valuable suggestions on enabling more open-ended interactions without scaffolding such as letting users enter their own explanations and hiding the test case categories hints. We discuss options on fading scaffolding in \cref{subsec:discuss-learning}.
% In particular, \emph{offloading code completion skills} for enabling interactive debugging seems promising, as 
Students also generally felt that the LLM-generated bugs and fixes were authentic. Most participants could not tell if their practiced programs were written by students or AI because of their experiences making or seeing similar mistakes from peers.

% When asked ``Do you think the buggy solutions you've seen are your fellow students' submissions or AI-generated? Why?'', most participants answered \texttt{Cannot tell} because of their experiences making or seeing similar mistakes from peers.
\begin{comment}
For example, students agreed with the statement that ``the fixed code reasonably reflects the explanation'' ($5.53 \pm 0.96$ out of 7).
eight participants answered \texttt{Cannot tell} because of their experiences making or seeing similar mistakes from peers (5) and not enough programming experiences (4).
For example, U11 mentioned \quoteinline{I wrote some code that was similar to these, and I saw some peers making these mistakes}.
Another eight participants mostly selected \texttt{AI's code} due to their weird logic (3), clean writing style (2), ease of fix (2), \quoteinline{just guessed} (U18), and \quoteinline{because you asked this question} (U1); three selected \texttt{Student's code} because \quoteinline{they look like genuine errors} (U13).
This echoes prior observations that LLMs can exhibit common mistakes similar to humans~\cite{fan2022automated}.
Interestingly, participants have different expectations of AI's code. For example, U6 believes that the readability of code is a sign that this can be written by a human student, while U10 believes that it is AI's code because \quoteinline{the indentations are always correct and the code is neat}.
\end{comment}

% \paragraphBold{\sysname's teachable agent setup supports students' engagement and confidence in debugging.}
Moreover, students reported that \sysname was engaging, fun, not frustrating, and helped build confidence in debugging.
% For the question ``How confident are you in terms of your debugging ability?'', 
A Wilcoxon signed-rank test shows a significant increase in self-rated confidence in debugging by 15\% ($p=0.007$). %from pre-survey to post-survey
% Additionally, we asked all users to rate the engagement, fun, and frustration aspects of conventional debugging practice they have had versus \sysname on a scale of 1 to 7.
Students rated \sysname as significantly more engaging (6.0 out of 7), fun (6.0), and less frustrating (2.5) than their conventional way of learning debugging and testing ($p < 0.005$ for each). % 4.1, 3.7, and 4.1 respectively, with 
% Participants commented that \sysname is \quoteinline{actually surprisingly fun}  (U1), \quoteinline{super cool} (U7), and \quoteinline{want to see it in the class} (U3). 
S8 especially liked the teachable agent setup: \quoteinline{the role play just feels more natural because it feels like explaining to a rubber duck instead of to talking to myself}.%, which reflects that LLM-based persona-mimicking could be promising for learning designs. 

\section{Discussion}
\label{sec:discuss}

\textbf{Teachable Agent for Appropriate Reliance with Imperfect AIs.}
\label{subsec:discuss-safe}
Our work illustrates a scenario in which \emph{LLM-generated bugs are not seen as problems but rather as features.}
\sysname's teachable agent setup provides students with \emph{moderated exposure to imperfect LLMs}, and may help them learn that LLMs are fallible and calibrate trust accordingly.
% build a realistic understanding of the fact that LLMs can generate bugs, before encountering them in uncontrolled, real-world scenarios.
Future iterations could remove material validation and allow direct exposure to unfiltered LLM mistakes in real-time interactions, taking full advantage of the teachable agent framework. Students will naturally expect that the LLM-agent seeking help may make mistakes (\eg fail to follow bug-fixing explanations).
This approach, however, requires a more sophisticated design for scaffolding students in recognizing LLM errors.

\textbf{Task Delegation for Shifting Learning Focus.}
\label{subsec:discuss-isolate}
Our exploration lays the foundation for a paradigm shift toward cultivating higher-order evaluation skills in the generative AI era.
Essentially, we asked: what skills should we offload, and what should we learn?
Most students in our study appreciated offloading subtasks to LLM (\cref{subsec:qualitative}); however, some need more scaffolds, while others prefer less. 
Future research can investigate more personalized task delegation. For example, students who need more help can use LLMs to facilitate code tracing, and students can also write their own explanations for bugs based on their proficiency.
Deciding the bare minimum programming skills and human-AI collaboration skills to teach also warrants further exploration~\cite{ma2023pair}.

\textbf{Modularize to Adapt to Different Needs.}
\label{subsec:discuss-material}
Though most students and instructors found \sysname engaging, 
some expressed concerns about the deployment and maintenance cost of a new tool.
To maximize utility to diverse users, we can modularize different components in \sysname. Instructors who prefer to distribute training materials as handouts can rely entirely on the material generation module. In contrast, instructors who want to experiment with TA training can employ \sysname with practice generated using their training questions. Future studies may perform ablation studies to evaluate different \sysname components with more extensive classroom deployment.

\textbf{Limitation.} We primarily evaluated \emph{whether \sysname can bring learning and efficiency gains} through small in-lab experiments. With this prerequisite, we plan to conduct future classroom deployment with controlled comparisons. There is also a limitation regarding the reported efficiency of the LLM-assisted instructional material development, as the instructors need some familiarization time with the tool and the process.

\section{Conclusion}
In an attempt to answer how LLMs can reshape programming education's focus, we introduce a novel system, \sysname, and new instructional designs for hypothesis construction skills. 
We aim to provide engaging and deliberate practice on debugging to novices, using our theoretically motivated and empirically tested teachable agent augmented by LLM.
Our evaluations show that \sysname can efficiently help instructors create high-quality instructional materials, effectively train novices on comprehensive and accurate hypothesis construction, and facilitate students' confidence and engagement in debugging. % Still, \sysname has room for improvement in scaffolding and interactivity. 

\subsubsection{Acknowledgments}
Thanks to the participants, reviewers, Vicky Zhou, Kelly Rivers, Michael Taylor, Michael Hilton, Michael Xieyang Liu, Kexin Yang, Jionghao Lin, Erik Harpstead, and other Ken's lab members for insights and help.
Thanks to the gift funds from Adobe, Oracle, and Google; Thanks to the National Science Foundation (award CNS-2213791) for partial support of this work.


\bibliographystyle{splncs04}
\bibliography{ref, paperpile}

% \appendix
% \input{sections/appendix}

\end{document}